\documentclass[
    ,final            
  ]
  {aipproc}

\layoutstyle{6x9}

\def\ls{{_<\atop^{\sim}}}
\def\gs{{_>\atop^{\sim}}}

\def\cm2{cm$^{-2}$}

\begin{document}

\title{Multiwavelength perspective of AGN evolution}

\classification{98.62.Js,87.59.-e,52.59.Px,98.54.-h}
\keywords      {galaxies: active  - galaxies: evolution -- multiwavelength surveys}

\author{Fabrizio Fiore}{
  address={Osservatorio Astronomico di Roma, via Frascati 33, Monteporzio, Italy}
}



\begin{abstract}
 
Discovering and studying obscured AGN at z$\gs1-3$ is important not
only to complete the AGN census, but also because they can pinpoint
galaxies where nuclear accretion and star-formation are coeval, and
mark the onset of AGN feedback. We present the latest results on the
characterization of z=1-3 galaxies selected for their high
mid-infrared to optical flux ratio, showing that they are massive and
strongly star-forming galaxies, and that many do host highly obscured
AGN. We present a pilot program to push the search of moderately
obscured AGN up to z=5-6 and discuss the perspectives of this line of
research.
\end{abstract}

\maketitle


\section{Introduction}

Understanding how galaxies formed and how they become the complex
systems we observe in the local Universe is a major theoretical and
observational effort, mainly pursued using multi-wavelength
surveys. We know today that tight links must exist between
super-massive black holes (SMBHs) found at the centre of bulgy
galaxies and their host galaxies. We also know that most SMBH growth
is due to accretion of matter during their active phases, implying
that most bulges went through a strong AGN phase to produce the
SMBH/galaxy mass ratio of $\sim0.001$ observed today.  The
co-evolution of galaxies and their SMBH depends on some physical
mechanism ('feedback' hereafter) linking accretion and ejection
occurring on sub-parsec scale in galaxy nuclei to the rest of the
galaxy. AGN feedback is often invoked to explain the observed galaxy
colors.  In the SMBH-galaxy co-evolutionary sequence the phase when a
galaxy is found in a passive status is preceded by a powerful active
phase, when cold gas is available for both star-formation and nuclear
accretion. The same cold gas and dust can intercept the line of sight
to the nucleus, and therefore a natural expectation of this scenario
is that the early, powerful AGN phase is also highly obscured. Once a
SMBH reaches masses $>10^{7-8}$M$_\odot$, the AGN can heat efficiently
the interstellar matter through winds, shocks, and high-energy
radiation, thus inhibiting further accretion and star-formation and
making the galaxy colors redder. At the end of this phase an
unobscured, type 1 AGN shines, while its host galaxy becomes
progressively red. Once most of the original cold gas is expelled from
the system, nuclear accretion and SF can occur only thanks to gas and
dust in stellar winds and/or accretion of fresh gas cooling from
haloes, and we are left with a passive or low star-forming galaxy,
hosting a relic SMBH or a low luminosity nucleus. While most known
systems are observed at the end of feedback processes, i.e. in the two
latter phases, the investigation of the coeval phase of obscured black
hole accretion and star-formation at z$\gs1-3$, the peak of nuclear
and stellar activity(\cite{ballantyne:2006,alex:2008,menci:2008} and
references therein), can give crucial information on AGN feedback when
it is in action. Current X-ray surveys can select efficiently
moderately obscured AGN up to z=2-3.  Highly obscured AGN at the same
redshifts can be recovered thanks to the dust reprocessing of the AGN
UV emission in the infrared, by selecting sources with mid-IR (and/or
radio) AGN luminosities but faint, host galaxy dominated, near-IR and
optical emission
\cite{martinez:2005,polletta:2006,daddi:2007,fiore:2008,dey:2008,fiore:2009}.
All these studies are based on Spitzer data, and were able to discover a
population of highly obscured AGN up to z$\sim3$. We summarize in Section 2
the latest results on this topic.  

Pushing this research up to the epoch of formation of the first
galaxies/SMBH at z$\gs3-4$ is clearly the next step. This may allow
one to assess the role of nuclear activity and AGN feedback in the
assembling of the first structures, and consequently to attack the
following fundamental problems.  First, optical surveys have
discovered luminous QSOs at z$>6$ with SMBHs of M$>10^9$M$_\odot$,
which would imply bulges of $>10^{12}$M$_\odot$ already formed less
than 1 Gyr after the Big Bang, if the local SMBH/galaxy mass ratio
should hold at those epochs. It is likely that the local SMBH/galaxy
relationship breaks down at some redshift, and indeed some indications
of this do exist\cite{alex:2008,peng:2006,lamastra:2010a}. But when?
and how? Second, early AGN activity can contribute to the reionization
and can heat the intergalactic medium, therefore affecting further
structure formation. Obtaining a complete AGN census at high-z is
crucial to tackle these issues. High-z AGN can however play a
crucial role also in other issues.  Structure formation and evolution
at high-z is naturally affected by the expansion rate of the Universe
at that epoch.  For example, a fast expansion rate at high-z may leave
too little time to form large structures, while a slow expansion rate
may imply a large amount of high-z galaxies and SMBHs. Therefore,
sensitive surveys of high-z galaxies and AGN may provide strong
constraints to the expansion rate of the Universe at early times. The
obvious problem is how to disentangle subtle cosmological effects from
complex baryon physics. The best strategy is to target quickly growing
structures, in particular structures growing exponentially, so that
little differences in the time of expansion of the Universe can be
significantly emphasized. SMBHs are the cosmic structures with the
fastest growth rate and this suggests using them to constrain AGN
feeding and accretion physics at high-z and to test and/or constrain
cosmological scenarios. We present in Section 3 preliminary results on
the search for high-z AGN in X-rays and discuss the perspectives of
this line of research.

\section{Probing the phase of SMBH/galaxy common growth: highly obscured AGN at z$\sim 1-3$}

\begin{figure}
\begin{tabular}{cc}
  \includegraphics[height=.25\textheight]{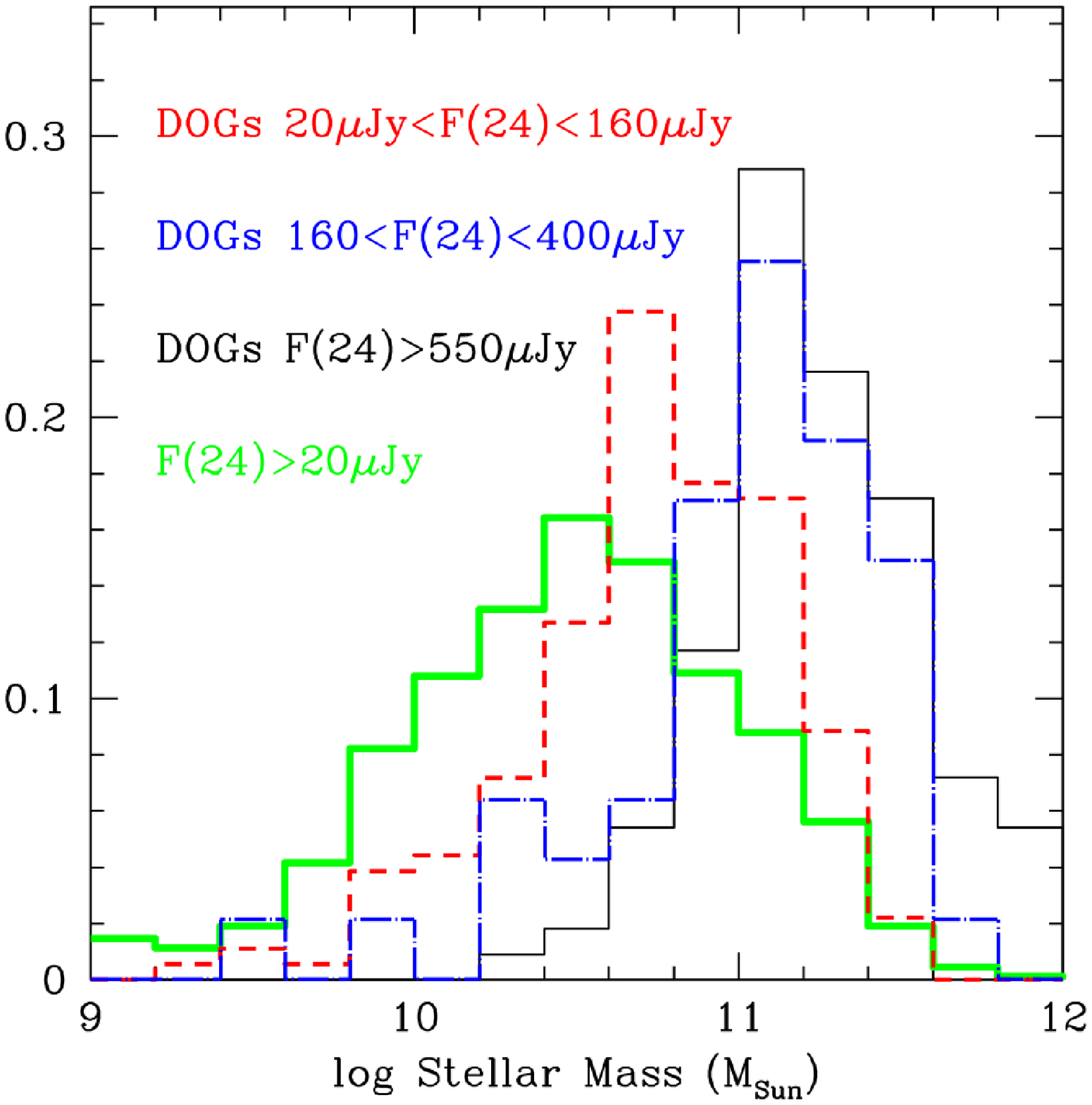} 
  \includegraphics[height=.25\textheight]{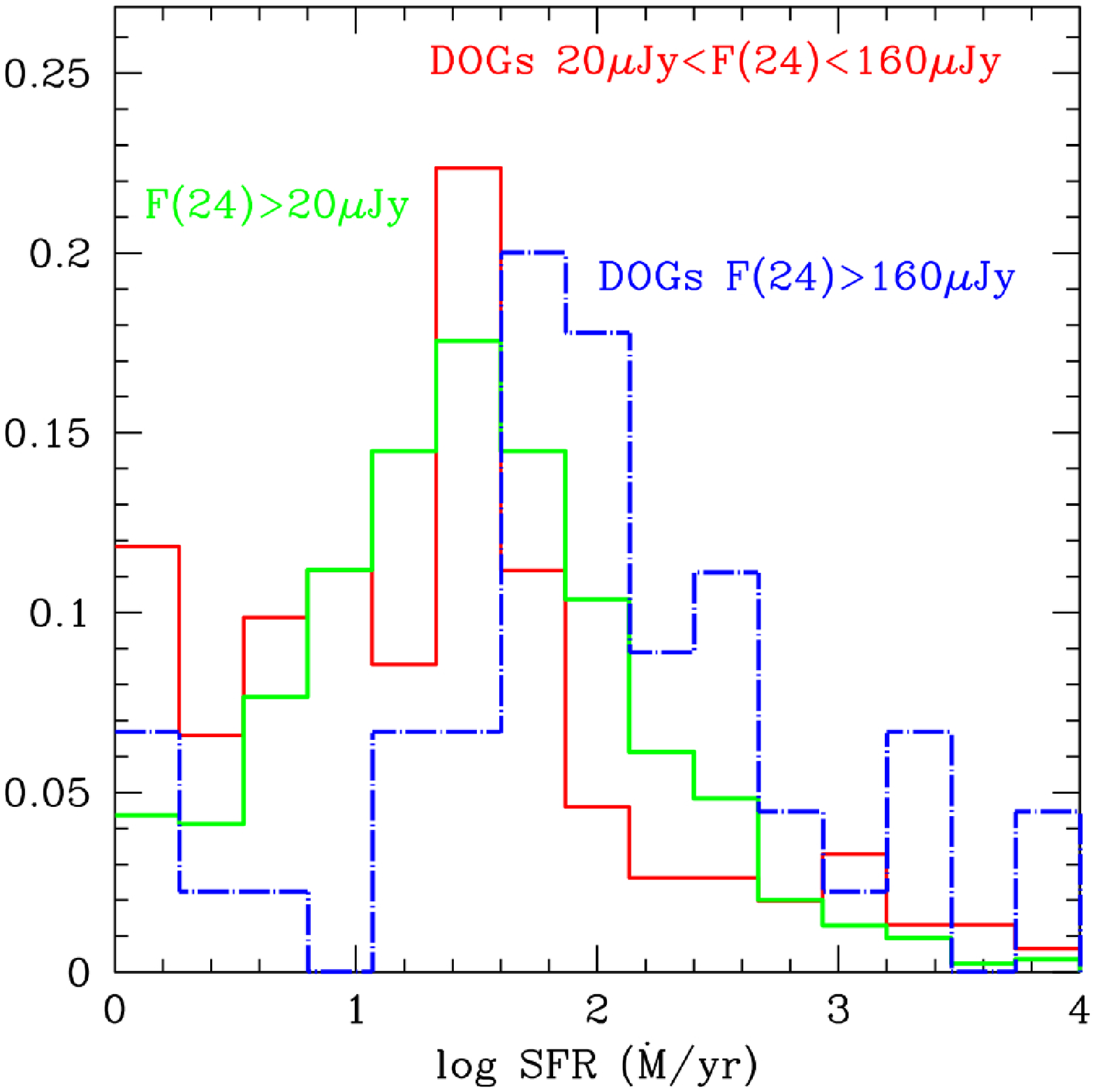}
\end{tabular}
\caption{Left panel: the stellar mass distribution of DOGs from the
COSMOS MIPS-bright sample (this solid line) and GOODS-MUSIC samples in
two F(24$\mu$m) bins: 160$<$F(24$\mu$m)$<400\mu$Jy (long dashed line);
20$<$F(24$\mu$m)$<160\mu$Jy (short dashed line). Right panel: the
star-formation rate distribution of GOODS-MUSIC DOGs with
F(24$\mu$m)$>160\mu$Jy (long dashed line); 20$<$F(24$\mu$m)$<160\mu$Jy
(thin solid line).  The stellar mass and star-formation rate (SFR)
distributions of the entire GOODS-MUSIC sample with
F(24$\mu$m)$>20\mu$Jy are shown for comparison in the respective panel
(thick solid lines).}
\end{figure}

Deep Spitzer surveys, accompanied by high quality multiwavelength
observations, have unveiled a large population of galaxies with high
mid-infrared to optical flux ratio (F(24$\mu$m)/F(R)$>$several
hundreds) and high redshift (z$\sim
1-3$, \cite{fiore:2008,dey:2008,fiore:2009}). These galaxies are often
dubbed Dust Obscured Galaxies or ``DOGs''. They are young, massive and
star-forming galaxies. Fig. 1, left panel, shows the stellar mass
distribution of DOGs from the COSMOS and GOODS-MUSIC samples
(F(24$\mu$m)/F(R)$>$300, R-K$>$4.5). The stellar mass has been
evaluated from the rest frame 1$\mu$m luminosity for the COSMOS
MIPS-bright sample (F(24$\mu$m)$>550\mu$Jy) and from a proper SED
fitting for the GOODS-MUSIC sample \cite{santini:2009}.  Fig. 1 right
panel shows the distribution of the star-formation rate for the
GOODS-MUSIC DOGs.  Note that the DOGs distributions are skewed towards
higher values of both stellar mass and SFR compared to the galaxy
population with F(24$\mu$m)$>20\mu$Jy.  The majority of DOGs with
F(24$\mu$m)$>160\mu$Jy have stellar mass $> 10^{11}M_{\odot}$ and
SFR$>40 M_{\odot}/yr$. Many DOGs have SFR of several hundreds or even
thousands $M_{\odot}/yr$. Similar results on masses and SFRs have been
recently obtained\cite{dey:2008,bussmann:2009}.  Several authors found
that a substantial fraction of DOGs hosts an active nucleus, based on
optical or mid-infrared
spectroscopy\cite{martinez:2005,houck:2005,weedman:2006,yan:2007}.
However, DOGs are usually faint X-ray sources in deep or even
ultra-deep X-ray observations. These facts suggest that DOGs often
host active nuclei that are highly obscured, nuclei covered by column
densities in excess to $\sim 10^{24}$cm$^{-2}$. Such high column
density strongly reduce the AGN hard X-ray flux, making them difficult
to study individually in X-rays. Nevertheless, DOGs are detectable in
deep Chandra surveys by stacking together the X-ray images at the
source position. This technique increases the exposure time by tens to
hundreds of times, and by a factor about square root of this number
the sensitivity and the flux limit. By using this technique we were
able to constrain the X-ray spectrum of DOGs in the GOODS-MUSIC and
COSMOS MIPS-bright samples, discovering hard X-ray colors. By
comparing these colors to the expectations of Monte Carlo simulations,
including spectral distributions of both star-forming galaxies and
obscured AGN, we constrained the highly obscured AGN fraction in the
COSMOS-MIPS bright DOGs with F(24$\mu$m)/F(R)$>$300 and $>1000$ to
$>60\%$ and $>90\%$ respectively\cite{fiore:2009}. Of course this is a
statistical results, valid on average for the sample, and valid if the
assumptions in the Monte Carlo simulations are correct. The most
critical assumption is the shape of the X-ray spectrum of purely
star-forming galaxies. We assumed, in agreement with X-ray
observations of nearby star-forming galaxies, a power law spectrum
with energy index $\alpha_E=0.9$ in the band 1-20 keV, which
corresponds to the observed band 0.3-6 keV at a typical redshift of 2.
In local galaxies the spectrum is dominated by the emission of low
mass X-ray binaries, which constitute about 80\% of the population of
luminous X-ray sources. High mass X-ray binaries, HMXB, have a
significantly flatter spectrum ($\alpha_E=0.2\pm0.2$) and their
fraction may well be higher in young star-forming
galaxies\cite{persic:2004}. In particular, their fraction may be
significantly higher if the galaxy IMF is skewed toward high stellar
mass values. We note however that the X-ray color measured for the
COSMOS MIPS sources with F(24$\mu$m)/F(R)$>$1000 and 300 corresponds
to an inverted spectrum with $\alpha_E=-0.5$ and to $\alpha_E=0$
respectively, quite extreme values even for galaxies with X-ray
emission completely dominated by HMXB. If DOGs do not host highly
obscured AGN their extremely hard X-ray color would put constraints on
their binary population, and therefore on the galaxy IMF.  The
cleanest way to identify a highly obscured AGN is through X-ray and/or
optical spectroscopy. The smocking gun of such AGN is a strong
(equivalent width EW$\gs 1$ keV) iron K$\alpha$ line. Furthermore,
strong high ionization lines such as [NeV] must also be present in the
optical spectra, together with the absence of broad lines. To
investigate further the nuclear properties of DOGs we have therefore
undertaken both lines of research.

\subsection{Rest frame X-ray spectroscopy of DOGs}

DOGs are faint X-ray sources and therefore direct X-ray spectroscopy
is feasible only for the brightest sources, i.e. DOGs in the SWIRE
survey\cite{polletta:2006,lanzuisi:2009}. Most CDFS and COSMOS DOGs
are too faint to extract useful X-ray spectral information from
individual sources. However, we may again resort to stacking
techniques to increase sample exposure time and sensitivity. To this
purpose we stacked together the X-ray counts of 99 COSMOS
MIPS-bright sources with F(24$\mu$m)/F(R)$>$300 and R-K>4.5 in 1 keV
wide rest frame bands, from 1 to 8 keV. We limited the analysis to
sources with less than 10 X-ray counts, to exclude unobscured and
moderately obscured AGN and to make sure that the stacks are not dominated
by a few bright sources. We also limited the analysis to sources in
the redshift range 0.7-3, for which the iron K$\alpha$ line is
observed at energies between 1.6 and 3.8 keV, with reasonably constant
efficiency.  Fig. 2 shows the stacked
Chandra images around the 99 COSMOS DOGs at 6 rest frame energies, and
the spectrum obtained by extracting the counts from circular regions
of 2 arcsec radius around the centers of these images. The DOGs are
detected with a signal to noise 3 and 3.5 in the 3-4 keV and 6-7 keV
bands respectively. Note as the signal is stronger in the 6-7 keV
band, which include the iron K$\alpha$ line, in comparison to the
nearby 5-6 keV and 7-8 keV bands (signal to noise 1.4 and 1.5
respectively). Since the width of the energy bins is 1 keV, only lines
with EW$\gs 1$keV would provide a signal. We conclude
that the detection of COSMOS DOGs at 6-7 keV and not at nearby
energies suggests the presence of strong K$\alpha$ lines
in a large fraction of these sources.

\begin{figure}
\begin{tabular}{cc}
  \includegraphics[height=0.2\textheight]{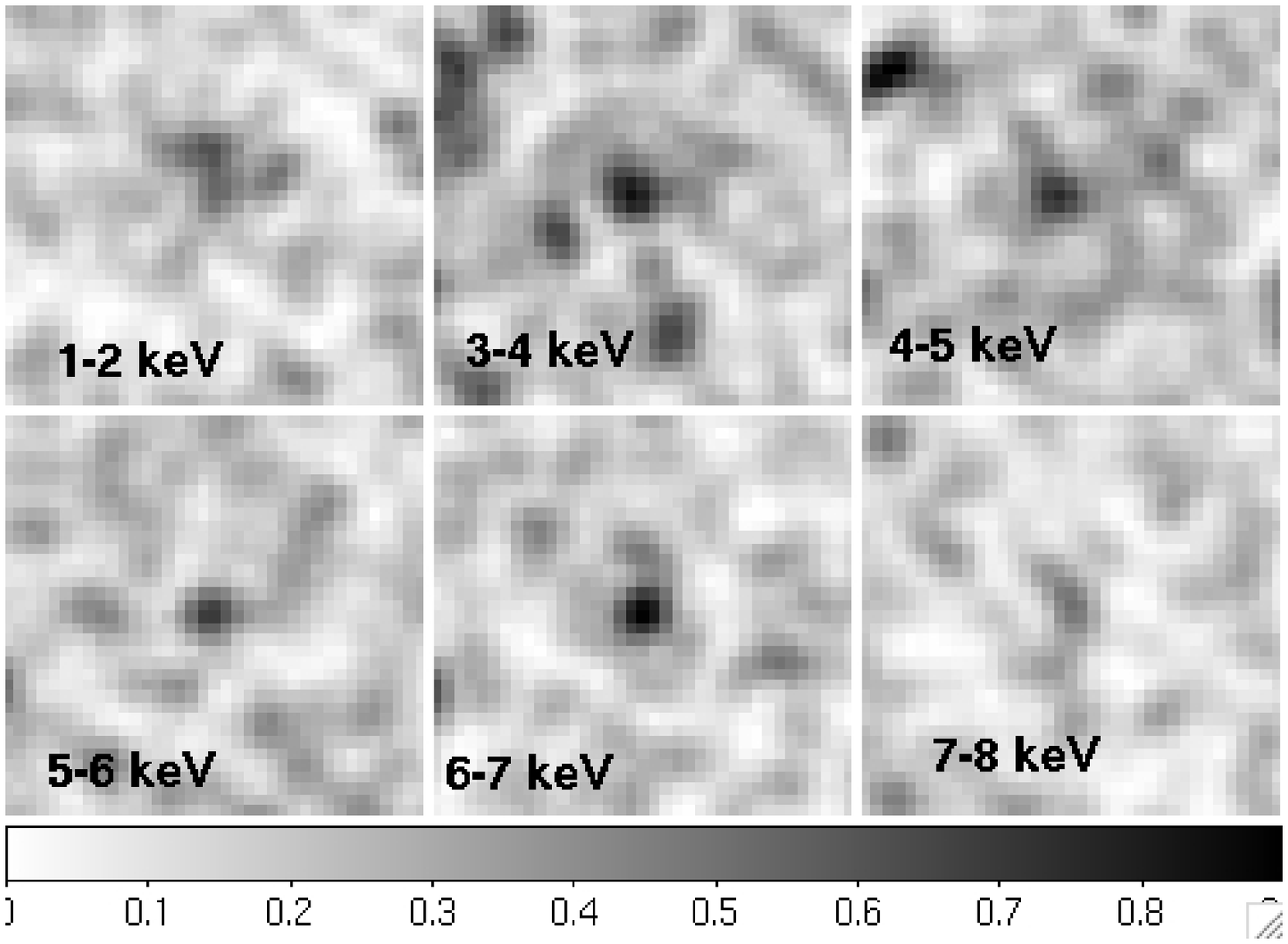}
  \includegraphics[height=.25\textheight]{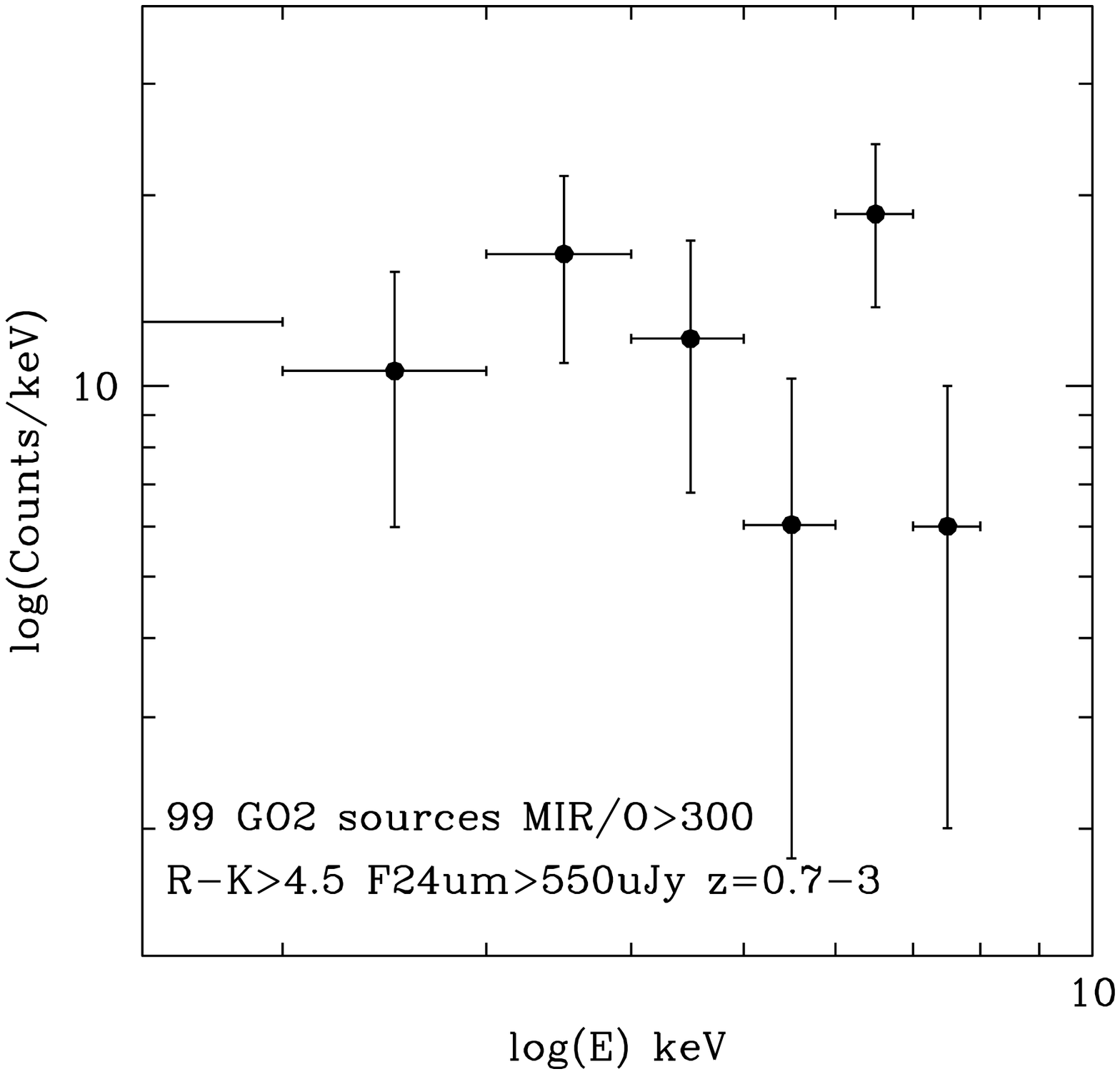}
\end{tabular}
  \caption{Left panel: Stacked Chandra images around the 99 COSMOS
DOGs at 6 rest frame energies. Right panel: Spectrum obtained by
extracting counts from the centers of images in left panel}
\end{figure}


\subsection{Optical spectroscopy of DOGs}

29 of the 171 COSMOS MIPS-bright sources with F(24$\mu$m)/F(R)$>$300
and R-K>4.5 have zCOSMOS and Magellan/IMACS optical spectra covering
the [NeV] line (but only 3 more extreme sources with
F(24$\mu$m)/F(R)$>$1000). We studied these spectra to search for high
ionization lines, indicative of AGN activity\cite{feruglio:2010}. We
detected [NeV] emission with EW=2-15\AA~ in 10 sources, 5 with
relatively bright X-ray counterparts in the Chandra-COSMOS
catalog\cite{civano:2010}, and therefore likely moderately obscured
AGN, and 5 in DOGs without a direct X-ray emission. The upper limits
to the [NeV] equivalent width are lower than 2-3\AA~ in 7 sources (5
without direct X-ray detection).  We conclude that $\gs50\%$ of the
COSMOS DOGs with F(24$\mu$m)/F(R)$>$300 and with optical spectra of
quality good enough to detect [NeV] do show AGN-like [NeV] emission, a
fraction consistent with the result of the Chandra stacking
analysis\cite{fiore:2009}.  We will more than double the number of
COSMOS DOGs with optical spectroscopy in the near future thanks to
additional zCOSMOS and IMACS spectra acquired in the course of
2008-2009, and to new Keck/DEIMOS spectroscopy currently on going.

\section{Pushing toward higher redshifts}

Joining current X-ray and infrared surveys will allow a sufficiently
robust determination of the luminosity function (LF) of all AGN,
regardless of the level of obscuration, and consequently the
determination of the full AGN census, up to z$\sim3$. At higher
redshifts the situation is much more complex.  The search of highly
obscured AGN at z$>3$ is today cumbersome, because of the limited
sensitivity of Spitzer for faint high-z sources. The selection of
large samples of highly obscured AGN at high z must probably await the
advent of JWST, with its factor of $\sim100$ improvement in
sensitivity in infrared imaging/spectrometry with respect to Spitzer,
and ALMA, with its spectroscopy capability. Thanks to the strong
negative K-correction at (sub-)mm wavelengths, ALMA may detect
star-forming galaxies at z$>$5 as easily as at z$\sim1$, regardless of
the presence of an AGN, and regardless of its obscuration. Hence ALMA
may deliver a truly unbiased sample of AGN.  The AGN identification
will be achieved through the detection of molecular lines (such as HCN
and HCO+), predominantly produced in "X-ray dominated regions" that
typically surround AGN\cite{maiolino:2008}.  On the other hand, large
area optical surveys such as the SDSS have already been able to
discover large samples of z$>4.5$ QSOs\cite{richards:2006} and about
40 QSOs at $z>5.7$\cite{jiang:2009}. The majority of the high-z SDSS AGN
are broad line, unobscured, high luminosity AGN. They are likely the
tips of the iceberg of the high-z AGN population.  Lower luminosity
and/or moderately obscured AGN can, in principle, be detected directly
in current X-ray surveys. Unfortunately, so far the number of X-ray
selected AGN at z$>3$ is only of a few tens, and $\sim$half a dozen at
z$>4.5$ (see
e.g. \cite{brandt:2004,fontanot:2007,brusa:2009a,brusa:2009b}). The
difficulty in detecting directly high-z AGN in X-ray surveys suggests
an alternative approach. Looking at the X-ray emission at the position
of known sources allows one to a) use a less conservative threshold for
source detection than in a blind search, and therefore to reach a
lower flux limit, and b) apply stacking
techniques\cite{lehmer:2005,aird:2008}.


As an experiment, we used the full 2 Ms exposure of the CDFS to look
at the X-ray emission at the position of $\sim 5000$ z$>2$ galaxies
drawn from the GOODS-MUSIC catalog\cite{grazian:2006}. To keep the
number of spurious X-ray detections small, we set the probability
threshold for a background fluctuation to $P=2\times10^{-4}$. First,
an average background spectrum was accumulated at three different
off-axis angles, by excluding bright sources\cite{luo:2009}.  The
inner background spectra show a stronger low energy cut-off, due to a
thicker deposition of contaminants at the centre of the ACIS-I 4 chip
region. Background at the position of each source was then evaluated
by normalizing the average background at the source off-axis to the
source spectrum accumulated above 7 keV, where the contribution of
faint sources is negligible, due to the sharp decrease of the mirror
effective area. We then computed the Poisson probability that the
counts accumulated below 7 keV are a fluctuation of this
background. We optimized the source detection both for the source
extraction area and for the energy band. We finally converted the
count rates in each detection band into 0.5-2 keV count rates by
assuming a power law model with $\alpha_E=0.4$. Assuming other
reasonable spectral shapes changes the conversion factor by
$\ls30\%$, thanks to the fact that the chosen reference band is
rather narrow. Using this technique we detected 33 z$>3$ galaxies in
the GOODS-MUSIC area, 15 of which are not present in previous catalogs
based on blind searches\cite{luo:2009} (10 galaxies with a
spectroscopic redshift). 5 galaxies have z$>4.5$ (1 spectroscopic
redshift).

\begin{figure}
\begin{tabular}{cc}
  \includegraphics[height=.23\textheight]{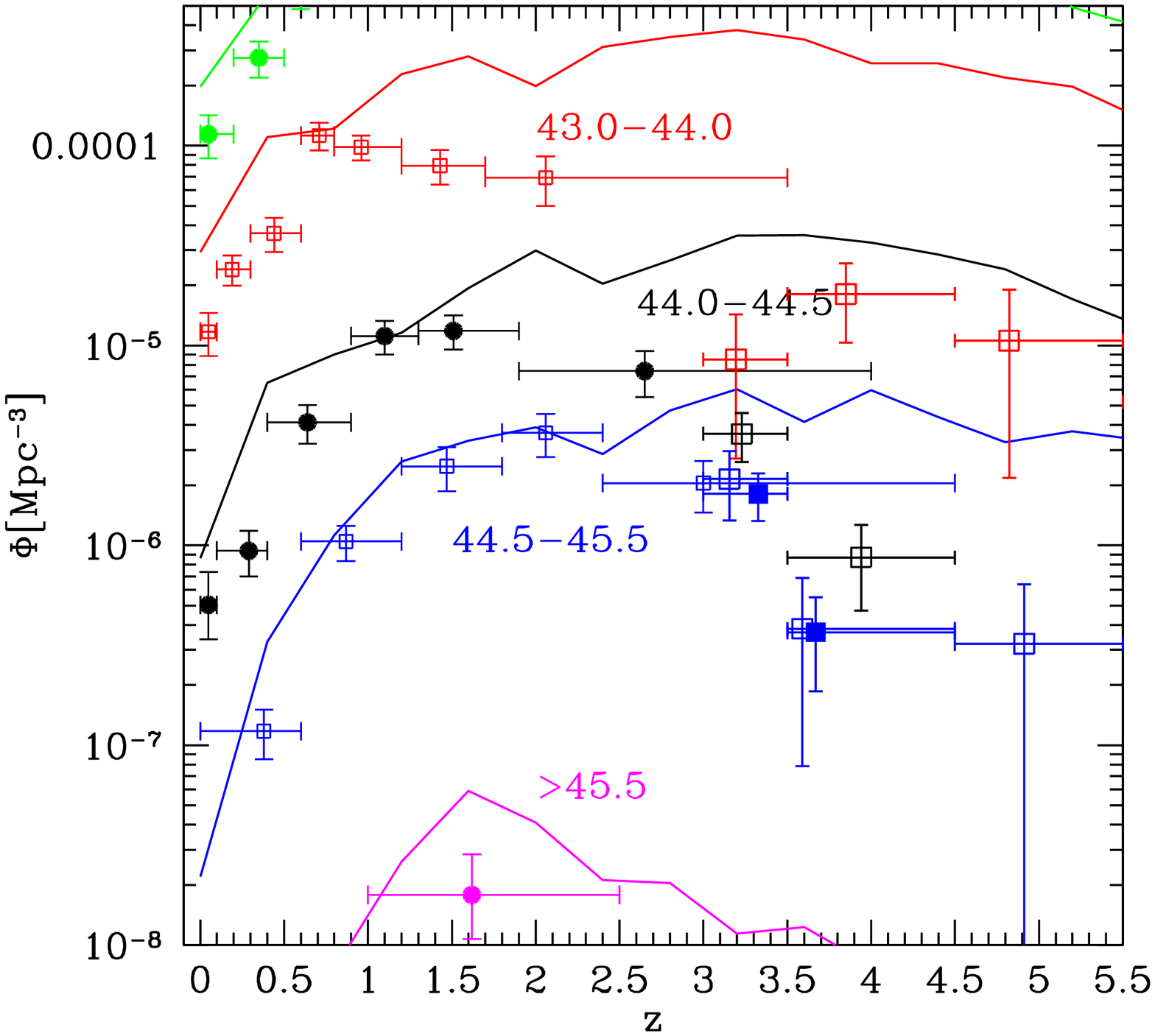}
  \includegraphics[height=.25\textheight]{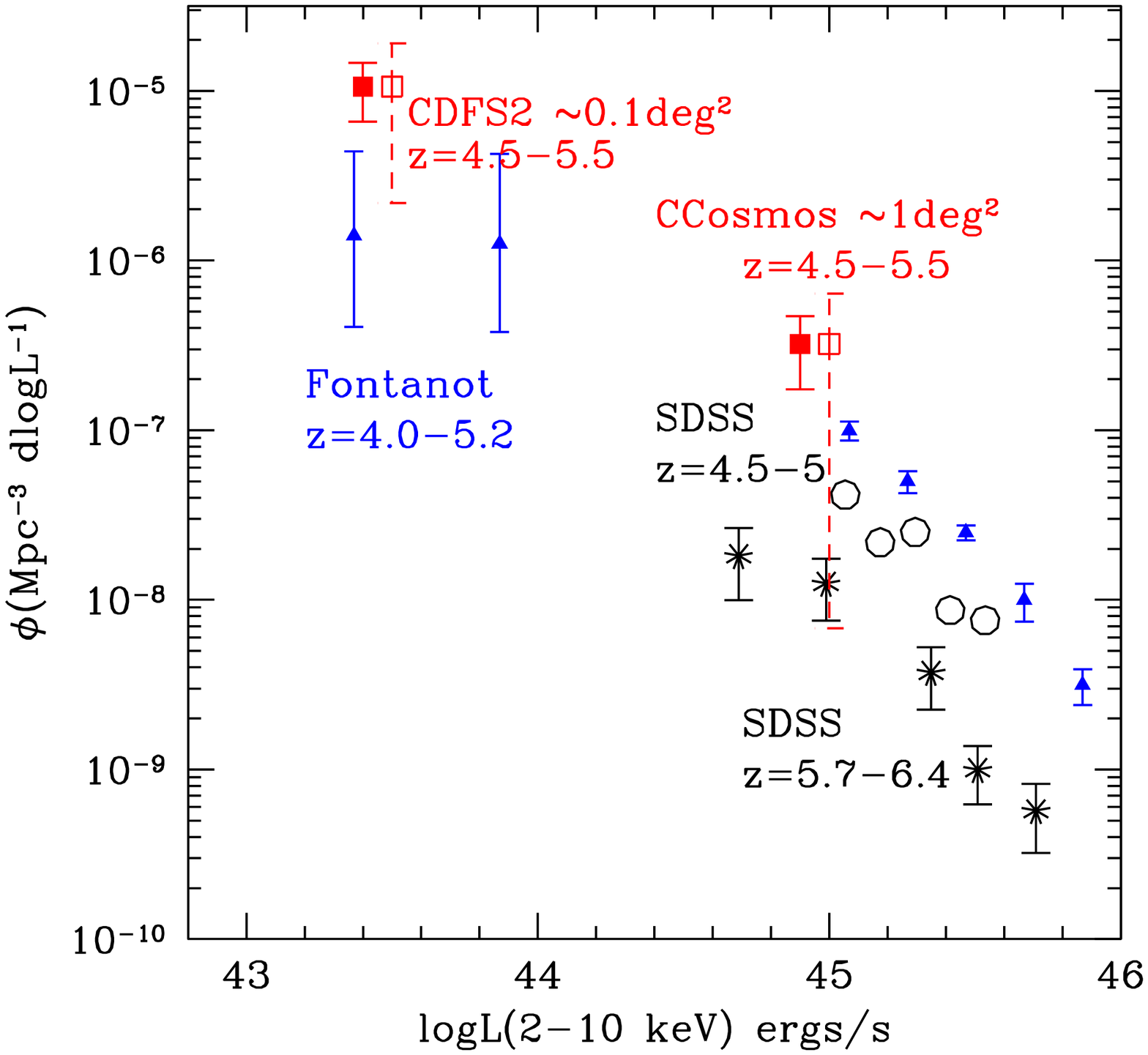}
\end{tabular}
  \caption{X-ray and optically selected high-z luminosity functions.
Open squares: this work; filled squares: same X-ray determination but
assuming 3 times more sources, e.g. about doubling the present CDFS
and COSMOS exposure times; stars: SDSS QSOs from Jiang et al. 2009;
open circles: SDSS QSOs from Richards et al. 2006; filled triangles:
high-z AGN from the Fontanot et al. 2007 compilation.}
\end{figure}

We combined this high-z sample to the Chandra-COSMOS
sample\cite{civano:2010} to compute the high-z AGN number density and
LF shown in Fig. 3. We plot in the same Fig. three optically selected
AGN LF, after converting the optical/UV luminosities to X-ray
luminosities following\cite{marconi:2004}. We note that first most of
the CDFS and COSMOS z$>3$ AGN are moderately obscured, based on a
hardness ratio analysis. They may well be missed in optical
surveys. Second, the X-ray LF extends the SDSS determination by $\sim2$
decades toward lower luminosities, reaching the regime of Seyfert-like
AGN at z=5. Third, our detemination at low luminosity is
nominally a factor $\sim 3-5$ higher than that of Fontanot et
al. (2007), although consistent with it within the large
error-bars. Reducing these error-bars is therefore mandatory. Doubling
the present CDFS and COSMOS exposure times would produce error-bars
comparable to those of optically selected surveys (Fig.3, right
panel).

\section{Conclusions}

Optical spectroscopy of COSMOS MIPS-bright DOGs with
F(24$\mu$m)/F(R)$>300$ and a rest frame stacking analysis of their
faint Chandra counterparts confirm that $\gs50\%$ do host a highly
obscured active nucleus, in agreement with\cite{fiore:2009}. DOGs are
massive, star-forming galaxies at z=1-3. They represent a class of
high-z galaxies where star-formation and nuclear accretion are coeval,
i.e. objects where AGN feedback is in action, and it has not yet
significantly quenched star-formation. Sub-millimiter galaxies share
the same characteristics\cite{alex:2008} and have similar space
densities\cite{alex:2008,fiore:2009}.  DOGs observations at mm
wavelength with present and future (ALMA) facilities may be able to
find the observational signature of outflows on galaxy scales, and
therefore of on going AGN feedback, through spectroscopy of molecular
lines.

We are starting to push the search for moderately and highly obscured
AGN up to the epoch of formation of the the first galaxies/SMBH,
i.e. up to z=6. Aggressive multiband data analysis strategies, fully
exploiting the synergies between optical/infrared and X-ray deep
surveys, can significantly increase the number of X-ray selected AGN
at high-z. Our pilot program on the CDFS 2 Msec observation nearly
doubled the number of $z>3$ AGN detected in X-rays in the GOODS-MUSIC
area.  Doubling the current Chandra exposure times on the CDFS and
COSMOS area will allow us to nearly triple the current sample of
high-z AGN in these surveys, which in turn will produce error bars
comparable to those of the z$>6$ QSO in the SDSS (see Fig. 3). A
program to double the Chandra exposure of the CDFS has been recently
approved by the CXO director, and these data will be available in the
next few years.  By joining the X-ray and SDSS selected high-z AGN
will allow us to constrain both the z$>4.5$ normalization of the AGN
LF and its slope over a wide luminosity interval (more than 3
decades).  At z=5-6 the SMBH mass growth is an extremely steep
function of the efficiency of conversion of gravitational potential
into radiation $\epsilon$. Since at such high-z it is reasonable to
assume that nuclear accretion proceeds at its Eddington value, this
means that also the AGN luminosity is a steep function of
$\epsilon$. This suggests that the slope of the AGN LF at high-z can
be used to constrain the $\epsilon$ and thus the black hole
spin\cite{volonteri:2005,king:2008}. Once fixed the BH spin
distribution through the measure of the slope of the AGN LF, the
normalization of the AGN LF may be used to constrain the expansion
rate of the Universe at early times, and therefore distinguish between competing
cosmological scenarios (Lamastra et al. in preparation).


\begin{theacknowledgments}
FF acknowledges support from ASI-INAF contract I/088/06/0 and PRIN-MUR
2006025203. This work would not had been possible without the help of
a large group a people including the GOODS-MUSIC and COSMOS teams.  I
would like to thank in particular S. Puccetti, C. Feruglio,
A. Lamastra, N. Menci, M. Brusa, F. Civano, A.  Comastri, C. Vignali,
M. Elvis, A. Fontana, P. Santini, M. Pirone, G. Lanzuisi, E. Piconcelli and M. Salvato.
\end{theacknowledgments}



\begin{thebibliography}{9}

\bibitem{aird:2008}
J. Aird et al. \emph{Montly Notices of the Royal Astronomy Society}, \textbf{387}, 883 (2008)

\bibitem{alex:2008}
D. Alexander et al. \emph{The Astronomical Journal}, \textbf{135}, 1968 (2008)

\bibitem{ballantyne:2006}
D.R. Ballantyne et al. \emph{The Astrophysical Journal}, \textbf{639}, 740 (2006)

\bibitem{brandt:2004}
W.N. Brandt et al. arXiv:0411355 (2004)

\bibitem{brusa:2009a}
M. Brusa et al. \emph{The Astrophysical Journal}, \textbf{693},
8 (2009)

\bibitem{brusa:2009b}
M. Brusa et al. \emph{Astronomy \& Astrophysics}, \textbf{507},
1277 (2009)

\bibitem{bussmann:2009}
R.S. Bussmann et al., \emph{The Astrophysical Journal}, \textbf{705},
184 (2009)

\bibitem{civano:2010}
F. Civano et al. in preparation (2010)

\bibitem{daddi:2007}
E. Daddi et al.  \emph{The Astrophysical Journal}, \textbf{670}, 173 (2007)

\bibitem{dey:2008}
A. Dey et al. \emph{The Astrophysical Journal}, \textbf{677},
943 (2008)

\bibitem{feruglio:2010}
C. Feruglio et al. in preparation (2010)

\bibitem{fiore:2008}
F. Fiore et al. \emph{The Astrophysical Journal}, \textbf{672},
94 (2008)

\bibitem{fiore:2009}
F. Fiore et al. \emph{The Astrophysical Journal}, \textbf{693},
447 (2009)

\bibitem{fontanot:2007}
F. Fontanot et al. \emph{Astronomy \& Astrophysics}, \textbf{461},
39 (2007)

\bibitem{grazian:2006}
A. Grazian et al. \emph{Astronomy \& Astrophysics}, \textbf{449},
951 (2006)

\bibitem{jiang:2009}
L. Jiang et al. \emph{The Astronomical Journal}, \textbf{138},
305 (2009)

\bibitem{houck:2005} 
J.R. Houck et al. \emph{The Astrophysical Journal}, \textbf{622}, L105 (2005) 

\bibitem{king:2008}
A. King et al. \emph{Montly Notices of the Royal Astronomy Society}, \textbf{385}, 1621 (2008)

\bibitem{lamastra:2010a}
A. Lamastra et al.  (2010) arXiv:1001.5407 (2010)

\bibitem{lanzuisi:2009} 
G. Lanzuisi et al. \emph{Astronomy \& Astrophysics}, \textbf{498}, 67 (2009) 

\bibitem{lehmer:2005}
B.D. Lehmer et al. \emph{The Astronomical Journal}, \textbf{129}, 1 (2005) 

\bibitem{luo:2009}
B. Luo et al.  \emph{The Astrophysical Journal}, \textbf{695}, 1227 (2009) 

\bibitem{maiolino:2008}
R. Maiolino arXiv:0806:0695 (2008)

\bibitem{marconi:2004}
A. Marconi et al. \emph{Montly Notices of the Royal Astronomy Society},
\textbf{351}, 169 (2004)

\bibitem{martinez:2005} 
A. Martinez-Sansigre et al.  \emph{Nature}, \textbf{436}, 666 (2005) 

\bibitem{menci:2008} 
N. Menci et al. \emph{The Astrophysical Journal}, \textbf{686}, 219 (2008)

\bibitem{peng:2006}
C.Y. Peng et al. \emph{The Astrophysical Journal}, \textbf{649}, 616 (2006)

\bibitem{persic:2004}
M. Persic et al. \emph{Astronomy \& Astrophysics}, \textbf{419}, 849 (2004)

\bibitem{polletta:2006} 
M. Polletta et al. \emph{The Astrophysical Journal}, \textbf{642}, 673 (2006)

\bibitem{richards:2006}
G.T. Richards et al. \emph{The Astronomica Journal}, \textbf{131}, 2766 (2006)

\bibitem{santini:2009}
P. Santini et al. \emph{Astronomy \& Astrophysics},  \textbf{504},
751 (2009)

\bibitem{volonteri:2005}
M. Volonteri, M. Rees \emph{The Astrophysical Journal}, \textbf{633}, 624 (2005)

\bibitem{weedman:2006} 
D.W. Weedman et al.   \emph{The Astrophysical Journal}, \textbf{651}, 101 (2006)

\bibitem{yan:2007} 
L. Yan \emph{The Astrophysical Journal}, \textbf{658}, 778 (2007)

\end{thebibliography}
\end{document}